\documentclass{article} 
\usepackage{iclr2026_conference,times}

\usepackage{amsmath,amsfonts,bm}

\def\eqref#1{equation~\ref{#1}}

\def\1{\bm{1}}

\DeclareMathAlphabet{\mathsfit}{\encodingdefault}{\sfdefault}{m}{sl}
\SetMathAlphabet{\mathsfit}{bold}{\encodingdefault}{\sfdefault}{bx}{n}

\usepackage{hyperref}
\usepackage{url}
\usepackage{graphicx} 

\usepackage{subcaption}
\usepackage{booktabs}

\title{When AI Agents Disagree Like Humans:\\Reasoning Trace Analysis for Human-AI\\Collaborative Moderation}

\author{Michał Wawer, Jarosław A. Chudziak \\
Faculty of Electronics and Information Technology\\
Warsaw University of Technology, Warsaw, Poland \\
\texttt{\{michal.wawer.stud,jaroslaw.chudziak\}@pw.edu.pl} \\
}

\iclrfinalcopy 
\begin{document}

\maketitle

\begin{abstract}
When LLM-based multi-agent systems disagree, current practice treats this as noise to be resolved through consensus. We propose it can be signal. We focus on hate speech moderation, a domain where judgments depend on cultural context and individual value weightings, producing high legitimate disagreement among human annotators. We hypothesize that convergent disagreement, where agents reason similarly but conclude differently, indicates genuine value pluralism that humans also struggle to resolve. Using the Measuring Hate Speech corpus, we embed reasoning traces from five perspective-differentiated agents and classify disagreement patterns using a four-category taxonomy based on reasoning similarity and conclusion agreement. 
We find that raw reasoning divergence weakly predicts human annotator conflict, but the structure of agent discord carries additional signal: cases where agents agree on a verdict show markedly lower human disagreement than cases where they do not, with large effect sizes ($d>0.8$) surviving correction for multiple comparisons. Our taxonomy-based ordering correlates with human disagreement patterns.
These preliminary findings motivate a shift from consensus-seeking to uncertainty-surfacing multi-agent design, where disagreement structure - not magnitude - guides when human judgment is needed.
\end{abstract}

\section{Introduction}
Content moderation presents a fundamental challenge for automated systems: decisions must be made at massive scale while navigating genuine disagreement about what constitutes harmful content \citep{gorwa2020algorithmic,moderationCaseStudy2023}. A comment referencing cultural practices may appear offensive to moderators unfamiliar with the context but appropriate to those within the community. These are not edge cases to be engineered away but endemic features of the moderation task \citep{hartmann2025lost,tomasev2021fairness,rieder2021fabrics}.
Multi-agent architectures have emerged as a promising approach to this challenge \citep{harbar2025simulating,zamojska2025games}, leveraging collective reasoning among multiple AI agents to improve decision quality \citep{du2024improving, chen2024reconcile}. Current implementations treat inter-agent disagreement primarily as a problem to overcome through consensus protocols \citep{zheng2025rethinkingreliabilitymultiagentsystem}, weighted voting \citep{jo2025byzantine,weightedbft2025} or Byzantine fault tolerance mechanisms that filter disagreeing agents as potentially unreliable \citep{chen2024blockagents}.
We argue this approach discards valuable information. When AI agents with similar capabilities persistently disagree about a judgment, this disagreement itself carries diagnostic meaning: it may signal that the content occupies a contested semantic region where human reasoners would also disagree. Rather than treating such disagreement as noise to be eliminated, we propose that AI disagreement can serve as a proxy for human cognitive disagreement, enabling systems to recognize when to defer to human judgment.

We connect collective intelligence research to multi-agent AI design, identifying conditions under which agent disagreement mirrors patterns of human cognitive disagreement. Our central hypothesis is that disagreement patterns in multi-agent deliberation can be systematically analyzed to distinguish cases requiring automated resolution from those requiring human judgment, transforming multi-agent moderation from a consensus-seeking system into an uncertainty-surfacing one. Based on it, we introduce a methodology for analyzing reasoning trace divergence that distinguishes value-based disagreement from error-based disagreement, paralleling how humans distinguish principled disagreement from mere confusion. Our initial results reveal a counterintuitive pattern: the magnitude of reasoning divergence between agents correlates negatively with human disagreement ($r = -0.19$), suggesting that how much agents diverge is uninformative. However, \emph{how} they diverge matters, our four-category taxonomy separates cases by human disagreement level (Kruskal-Wallis $H = 54.0$), with the agreement-disagreement boundary driving the effect. This motivates a shift from measuring divergence to characterizing its structure.

\section{Theoretical Framework}

Human disagreement about evaluative judgments often reflects genuine value pluralism rather than error \citep{gajewska2026,mlBasedContentModerationSystems2022}. Cognitive science research demonstrates that humans weigh competing values differently when reasoning about complex social situations \citep{Liao2011,sadowskiLegalReasoning2025}. Whether strong political speech constitutes harassment depends on how one weighs free expression against community protection and different people applying consistent principles may disagree. In AI systems outputs, rather than assuming a ground truth that disagreeing agents are failing to reach consensus, we can treat certain disagreement patterns as informative signals that content involves genuine value tensions \citep{wang2025impscore,wang2025mixtureofagents}.

Recent work confirms these concerns. Studies of multi-agent debate find that majority pressure can suppress independent correction \citep{huang2025resilience}, with extended debate rounds sometimes entrenching errors rather than correcting them \citep{liang2024encouraging}. Evaluations across multiple debate frameworks find they ``fail to consistently outperform simpler single-agent strategies'' \citep{zhang2025multillmagents} suggesting multi-agent deliberation may degrade to inefficient resampling without genuine epistemic diversity \citep{du2024improving}.

We propose distinguishing two sources of persistent disagreement, inspired by cognitive accounts of human reasoning: (1) Value pluralism (convergent disagreement): Agents reasoning from different implicit value weightings reach different conclusions through valid reasoning chains. When agents understand content similarly but judge it differently, the disagreement reflects legitimate value tensions, mirroring how humans with shared understanding can reach different evaluative conclusions. (2) Noise or error (divergent disagreement): Agents make inconsistent judgments due to different interpretations, context gaps or reasoning failures \citep{kostka2025synergizinglogicalreasoningknowledge}. This parallels cases where human disagreement stems from miscommunication rather than genuine value differences. The methodological challenge is distinguishing these sources. We propose that analyzing the structure of disagreement, not just its presence, enables better classification for content moderation.

Our approach connects to formal results in social choice and collective intelligence. The Condorcet Jury Theorem \citep{austen1996information} predicts that majority voting
among independent agents converges to correct decisions, but recent work shows LLM agents violate the independence assumption due to shared pretraining, causing consensus without
correctness gains \citep{denisovblanch2025consensus}. This failure mode motivates our shift from consensus-seeking to disagreement analysis. More broadly, judgment aggregation
theory \citep{list2002aggregating} demonstrates that aggregating premise-level and conclusion-level votes can yield contradictory collective judgments, the discursive dilemma. Our taxonomy operationalizes a related insight: agents may agree on premises (high reasoning similarity) yet disagree on conclusions, and this structural pattern carries diagnostic value that raw aggregation discards.

\section{Methodology}

\begin{figure}[t]
\centering
\includegraphics[width=0.80\linewidth]{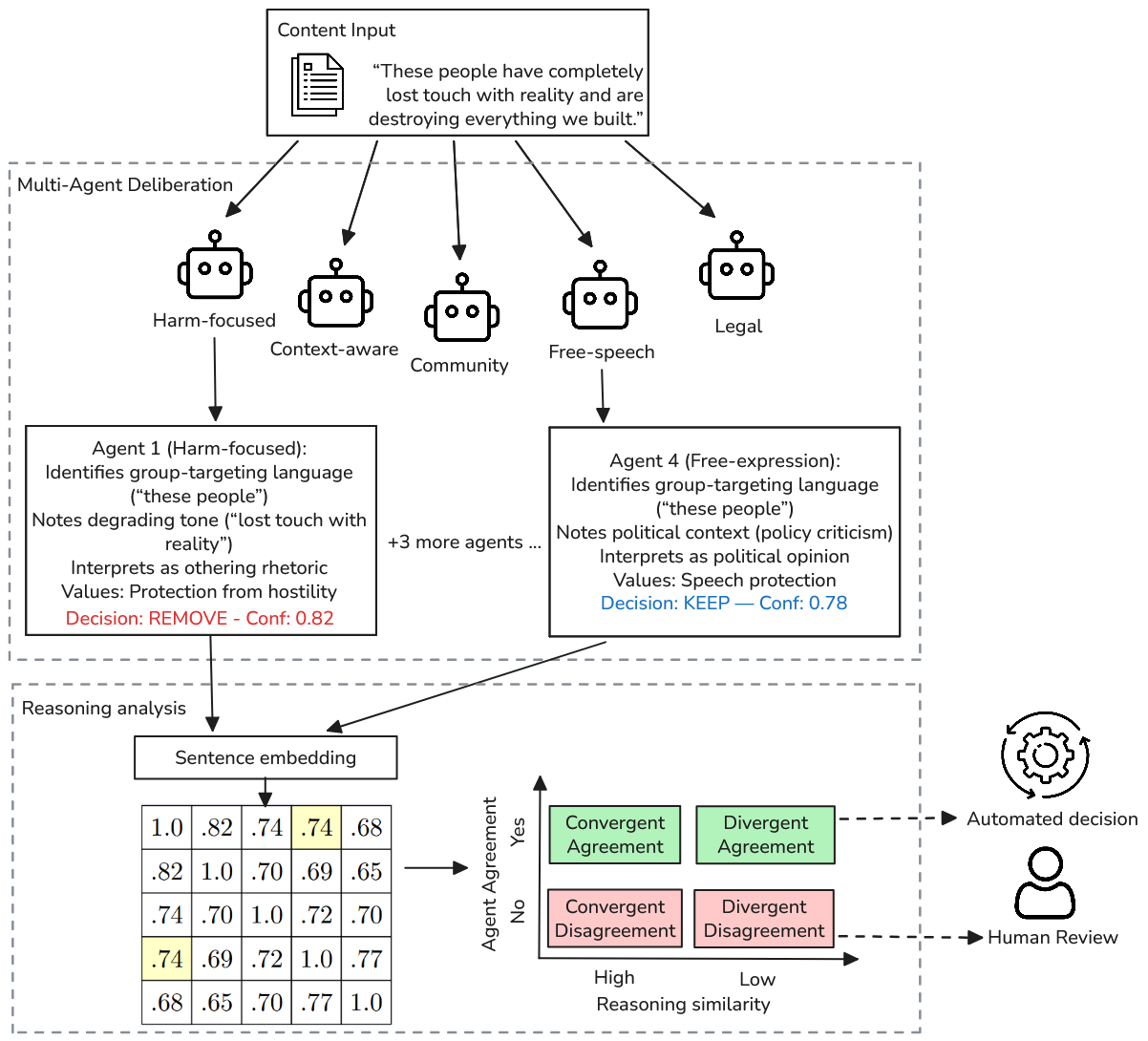}
\caption{Overview of our framework design, content is processed by $N$ perspective-differentiated agents, each generating reasoning traces. Traces are embedded and compared via cosine similarity. The disagreement taxonomy routes cases to automated resolution or human escalation.}
\label{fig:architecture}
\end{figure}

We employ a deliberation architecture with $N=5$ agents instantiated from the same base model (DeepSeek-V3) but differentiated through system prompts encoding distinct moderator perspectives. Each perspective emphasizes different values: harm prevention,
contextual interpretation, community norms, free expression,
and legal standards, respectively.

Design shown on Fig. \ref{fig:architecture} creates simulated perspective diversity while controlling for capability differences, all agents share the same underlying knowledge and reasoning capacity, differing only in value emphasis and framing. The distinctive vocabularies enable semantic differentiation of reasoning traces. For each content item, agents independently generate moderation decisions with explicit reasoning traces. We invoke structured justifications through chain-of-thought prompting \citep{wei2022chain}, requiring agents to articulate: (1) content interpretation, (2) relevant considerations, (3) value trade-offs, and (4) final judgment with confidence. Agents produce binary decisions (REMOVE/KEEP) with confidence scores and explicit statements of which values drove their judgment.

Reasoning traces are the intermediate justifications agents generate when arriving at decisions. Our methodology for analyzing these traces builds on \citet{tajik2026disagreement}, who introduced reasoning trace analytics with categories
analogous to ours (within-align/within-misalign). We extend their framework to content moderation, where we interpret convergent disagreement as a signal of value pluralism rather than a generic misalignment indicator. We embed traces into a shared semantic space for quantitative comparison. For each agent $i$ and content item $c$, let $r_i^c$ denote the full reasoning trace. We compute sentence embeddings using a pre-trained transformer model (all-mpnet-base-v2), yielding vector representations $\mathbf{e}_i^c \in \mathbb{R}^d$. We then compute pairwise cosine similarity:

\begin{equation}
    s_{ij}^c = \frac{\mathbf{e}_i^c \cdot \mathbf{e}_j^c}{\|\mathbf{e}_i^c\| \|\mathbf{e}_j^c\|}
\end{equation}

From pairwise similarities, we derive aggregate divergence metrics for each content item:

\begin{itemize}
    \item {Mean similarity}: $\bar{s}^c = \frac{2}{N(N-1)} \sum_{i<j} s_{ij}^c$
    \item {Minimum similarity}: $s_{\min}^c = \min_{i<j} s_{ij}^c$
\end{itemize}

Low mean similarity indicates agents are reasoning about the content in fundamentally different ways; low minimum similarity indicates at least one agent pair has highly divergent reasoning. We classify each content item into one of four categories based on the joint distribution of reasoning similarity and conclusion agreement:
\begin{enumerate}
\item {Convergent Agreement} (high similarity, same conclusion): Agents reason similarly and agree, representing confident decisions where collective intelligence operates normally.

\item {Divergent Agreement} (low similarity, same conclusion): Agents reason differently but converge on the same judgment through independent reasoning paths.

\item {Convergent Disagreement} (high similarity, different conclusions): Agents reason similarly but reach different judgments, we hypothesize this is strong signal of value pluralism.

\item {Divergent Disagreement} (low similarity, different conclusions): Agents reason and conclude differently, representing genuine edge cases or noise from inconsistent interpretation.
\end{enumerate}

\section{Experimental Design}

We use the Measuring Hate Speech corpus \citep{kennedy2020constructing, sachdeva2022measuring}, which aggregates annotations rating 39,565 social media comments.
Annotators rated each comment on 10 ordinal dimensions, using 5-point scales.
We compute human annotator disagreement for each content item as the standard deviation of ratings across annotators. Items with high standard deviation represent cases where human perspectives genuinely diverge; items with low standard deviation represent cases of relative consensus. As an initial validation of the proposed framework, we sample $n=600$ content items stratified by human disagreement level, with equal representation across three bins (200 low, 200 medium, 200 high disagreement). For each item, we:

\begin{enumerate}
    \item Generate moderation decisions from 5 perspective-differentiated agents via DeepSeek API (temperature $= 0.7$), collecting full reasoning traces
    \item Compute pairwise reasoning trace similarity using embeddings (all-mpnet-base-v2)
    \item Calculate aggregate divergence as $1 - \bar{s}^c$ (mean pairwise dissimilarity)
    \item Classify each item into the four-category taxonomy using similarity threshold $\theta_s = 0.72$ (set at approximately one standard deviation above the observed mean pairwise similarity of $0.67$) and majority agreement threshold $> 50\%$
    \item Record agent conclusions (binary REMOVE/KEEP decisions)
\end{enumerate}

We evaluate if raw agent reasoning divergence predicts human disagreement by computing Pearson and Spearman correlations between divergence scores and human annotator rating standard deviation. We then examine whether the structure of disagreement carries additional signal by comparing mean human disagreement across the four taxonomy categories and testing between-category differences with Kruskal-Wallis tests. We assess the practical value of the taxonomy for routing decisions by treating high human disagreement as ground truth and comparing category-based escalation against divergence-only and random baselines using precision, recall, and F1, where a true positive occurs when a predictor correctly flags a case that human annotators found genuinely contested.

\section{Preliminary Results}

Table \ref{tab:categories} summarizes the distribution of items across our four-category taxonomy and the mean human annotator disagreement within each category.

\begin{table}[h]
    \centering
    \small
    \begin{tabular}{lccc}
    \hline
    \textbf{Category} & \textbf{n} & \textbf{\%} & \textbf{Mean $d$} \\
    \hline
    Convergent Disagreement (CD) & 85 & 14.2 & 0.813 \\
    Divergent Disagreement (DD) & 361 & 60.2 & 0.763 \\
    Convergent Agreement (CA) & 23 & 3.8 & 0.666 \\
    Divergent Agreement (DA) & 131 & 21.8 & 0.356 \\
    \hline
    \end{tabular}
    \caption{Human disagreement ($d$) by agent disagreement category.}
    \label{tab:categories}
\end{table}

Agent reasoning divergence shows a weak negative correlation with human annotator disagreement (Pearson $r = -0.19$). The overall category effect is significant
(Kruskal-Wallis $H = 54.0$). Post-hoc pairwise comparisons reveal that this effect
is primarily driven by divergent agreement (DA) cases showing significantly lower human disagreement than both convergent disagreement (Cohen's $d = 0.89$) and divergent disagreement ($d = 0.80$).
The difference between CD ($\bar{d} = 0.813$) and DD ($\bar{d} = 0.763$) is not statistically significant, indicating that the primary diagnostic value of the taxonomy lies in distinguishing agreement from disagreement categories rather than differentiating
within disagreement types. We tested sensitivity to $\theta_s$ across $[0.60, 0.84]$: the Spearman correlation remains significant at all thresholds ($\rho \in [0.25, 0.30]$ in all cases), confirming that results are robust to the threshold choice.

\begin{table}[h]
    \centering
    \small
    \begin{tabular}{lcccc}
    \hline
    \textbf{Predictor} & \textbf{Precision} & \textbf{Recall} & \textbf{F1} \\
    \hline
    Category-based escalation & 0.401 & 0.845 & 0.548 \\
    Divergence Only & 0.347 & 0.915 & 0.503  \\
    Random Baseline & 0.333 & 0.505 & 0.401\\
    \hline
    \end{tabular}
    \caption{Predictor comparison for flagging high human disagreement cases. The escalation score achieves the highest F1; all informed methods outperform random assignment.}
    \label{tab:predictors}
\end{table}

The category-based escalation score achieves the highest F1 ($0.54$), compared to the divergence-only baseline ($0.50$) and random assignment ($0.40$). The improvement over divergence-only is modest and may not be significant at this sample size. Notably, the divergence-only predictor achieves higher recall ($0.915$ vs. $0.845$).
In safety-critical deployment, where missing a genuinely contested case is costlier than over-escalating, higher recall may be preferable. The taxonomy's advantage is therefore primarily \emph{diagnostic}, it provides interpretable categories explaining \emph{why} escalation is warranted, rather than purely predictive.

\section{Discussion and Future Work}

Our results suggest that multi-agent systems can implicitly model aspects of human cognitive disagreement, not through the magnitude of their divergence, but through its structure. The weak negative correlation of raw divergence ($r = -0.19$) compared with the positive taxonomy-based correlation ($\rho = 0.27$, escalation F1 $= 0.54$) provides preliminary evidence that \emph{how} agents disagree carries more diagnostic value than \emph{how much}. Both effects are modest, and the CD-DD distinction does not reach statistical significance at the current sample size. The significant finding is the separation between agreement categories (DA, CA) and disagreement categories (DD, CD).
The four-category taxonomy currently reduces in practice to a binary: agent agreement vs.\ disagreement. The full 2$\times$2 structure remains theoretically motivated-convergent
disagreement is where value pluralism should be most visible, but confirming this requires larger samples or architecturally diverse agents. The practical implication is a shift in design goals from optimizing for consensus toward optimizing for appropriate routing, where agent disagreement triggers human escalation rather than being suppressed. Our evaluation on 600 items from a single corpus establishes that structural analysis of disagreement shows promise, but substantial further validation is needed. The imbalanced category distribution - DD comprising 60.2\% of items, likely reflects our single-model design: because all agents share the same base model and differ only in system prompts, the perspective-specific vocabulary creates systematic semantic divergence even when agents reason substantively similarly. Architecturally diverse agents would likely produce a more balanced distribution by decoupling vocabulary from reasoning effects. Additional limitations include the focus on English-language U.S.\ social media and the absence of task-level evaluation. We plan to address both through expanded evaluation across multiple corpora and a study assessing whether taxonomy-based escalation improves moderation quality in practice.

\section{Conclusion}

We proposed treating multi-agent disagreement as diagnostic signal rather than noise in content moderation. By analyzing reasoning trace divergence through a four-category taxonomy, we can distinguish cases likely reflecting genuine value pluralism from those reflecting error - and route them accordingly. This reframes multi-agent moderation from consensus-seeking to collaborative sense-making, where disagreement structure guides the human-AI division of cognitive labor.



\bibliography{iclr2026_conference}
\bibliographystyle{iclr2026_conference}


\end{document}